\documentclass[aps,prl,reprint]{revtex4-2}
\usepackage{amsmath}
\usepackage{amssymb}
\usepackage{graphicx}
\usepackage{hyperref}
\usepackage{physics}
\usepackage{times}
\usepackage{xcolor}
\hypersetup{colorlinks, linkcolor={blue}, citecolor={blue}, urlcolor={blue}}

\renewcommand{\i}{\mathrm{i}}
\renewcommand{\a}{\mathbf{a}}
\renewcommand{\b}{\mathbf{b}}

\newcommand{\subref}[2]{\hyperref[#1]{\ref*{#1}(#2)}}

\begin{document}

\title{Observation of mHz-level cooperative Lamb shifts in an optical atomic clock}

\author{Ross B. Hutson}
\email{ross.hutson@colorado.edu}
\affiliation{JILA, NIST and University of Colorado, 440 UCB, Boulder, Colorado 80309, USA}
\affiliation{Department of Physics, University of Colorado, 390 UCB, Boulder, CO 80309, USA}
\author{William R. Milner}
\affiliation{JILA, NIST and University of Colorado, 440 UCB, Boulder, Colorado 80309, USA}
\affiliation{Department of Physics, University of Colorado, 390 UCB, Boulder, CO 80309, USA}
\author{Lingfeng Yan}
\affiliation{JILA, NIST and University of Colorado, 440 UCB, Boulder, Colorado 80309, USA}
\affiliation{Department of Physics, University of Colorado, 390 UCB, Boulder, CO 80309, USA}
\author{Jun Ye}
\email{jun.ye@colorado.edu}
\affiliation{JILA, NIST and University of Colorado, 440 UCB, Boulder, Colorado 80309, USA}
\affiliation{Department of Physics, University of Colorado, 390 UCB, Boulder, CO 80309, USA}
\author{Christian Sanner}
\affiliation{Department of Physics, Colorado State University, Fort Collins, Colorado 80523, USA}

\date{\today}

\begin{abstract}
We report on the direct observation of resonant electric dipole-dipole interactions in a cubic array of atoms in the many-excitation limit.
The interactions, mediated by single-atom couplings to the shared electromagnetic vacuum, are shown to produce spatially-dependent cooperative Lamb shifts when spectroscopically interrogating the mHz-wide optical clock transition in strontium-87.
We show that the ensemble-averaged shifts can be suppressed below the level of evaluated systematic uncertainties for state-of-the-art optical atomic clocks.
Additionally, we demonstrate that excitation of the atomic dipoles near a Bragg angle can enhance these effects by nearly an order of magnitude compared to non-resonant geometries.
Given the remarkable precision of frequency measurements and the high accuracy of the modeled response, our work demonstrates that such a clock is a novel platform for studies of the quantum many-body physics of spins with long-range interactions mediated by propagating photons.
\end{abstract}
\maketitle

{\em Introduction.---}Studies of quantum many-body physics naturally arise in the context of quantum sensing.
For any quantum sensor, the amount of extractable information regarding a metrological quantity of interest is fundamentally limited by the number of accessible qubits~\cite{Braunstein1994,Giovannetti2006}. 
This creates a generic incentive to build devices capable of manipulating and characterizing quantum systems of ever-increasing size~\cite{Caves1980,Itano1993,Scully1993}.
Since interactions within the system or with the environment typically scale with system size, the main challenges are then twofold: how can interactions be controlled to reduce systematic effects, and/or how can they be leveraged to generate useful entanglement?

In the context of atomic clocks, significant progress toward probing larger numbers of atoms, while avoiding systematic effects due to contact interactions, has been made by trapping atoms in three-dimensional optical lattices with at most one atom per lattice site~\cite{Akatsuka2010,Campbell2017,Marti2018,Oelker2019}.
Nonetheless, long-range interactions in the form of resonant dipole-dipole interactions have loomed just beyond experimental detectability~\cite{Chang2004,Kramer2016,Cidrim2021}.
In this letter we describe an experimental apparatus where such interactions can be precisely measured and controlled, demonstrating an understanding of how they can affect current and future generations of optical atomic clocks, or alternatively, how atomic clocks may offer insight into the nature of light-matter interactions. 
Beyond defining a novel platform for the study of effective photon-photon interactions~\cite{Asenjo-Garcia2017,Henriet2019}, such engineered arrays of narrow-band quantum emitters provide a path to a new class of photonic devices based on controlled collective atom-photon dynamics.

{\em Resonant dipole-dipole interactions.---}The classical electric field, evaluated at a position $\b$, generated by a point dipole $\mathbf{d}_\a\propto e^{-\i \omega t}$, oscillating at an angular frequency $\omega$, and localized at a position $\a$, is given by 
$\mathbf{E}_\a(\b) = k^3 e^{\i k r} \{[\mathbf{d}_\a - \hat{\mathbf{r}}(\hat{\mathbf{r}} \cdot \mathbf{d}_\a)] / (kr) + [3\hat{\mathbf{r}}(\hat{\mathbf{r}}\cdot\mathbf{d}_\a) - \mathbf{d}_\a][1 / (kr)^{3} - \i / (kr)^{2}]\} / 4\pi\epsilon_0 $,  where $\mathbf{r} = \mathbf{r}_\mathbf{ba} = \b - \a$, $r = |\mathbf{r}|$, $\hat{\mathbf{r}} = \mathbf{r} / r$, $\epsilon_0$ is the vacuum permittivity, and $k = \omega / c$ with $c$ being the speed of light~\cite{Jackson1998}.
A second, freely oscillating dipole $\mathbf{d}_\b$ localized at position $\b$ will then dynamically evolve according to the interaction term $H_\mathbf{ba} = -\mathbf{d}_\b \cdot \mathbf{E}^\ast_\a(\b)$ whose real and imaginary parts respectively lead to a frequency shift and damping of the initial excitation.
These interactions form the basis of classical linear optics~\cite{Jackson1998,deVries1998,Andreoli2021}.

An ensemble of indistinguishable (pseudo-)spin-$\frac{1}{2}$ systems, with internal ground and excited states labeled $\ket{g}$ and $\ket{e}$ respectively, can analogously be described using the formalism of quantum optics where the reduced density matrix $\hat{\rho}$ evolves in time according to the master equation $\partial_t \hat{\rho} = \mathcal{L}_\mathrm{free}[\hat{\rho}] = \mathcal{L}_1[\hat{\rho}] + \mathcal{L}_2[\hat{\rho}]$ with the Liouvillian superoperator describing collective electromagnetic interactions given in Lindblad form as~\cite{Lehmberg1970} 
\begin{equation}
\begin{aligned}
    \label{eqn:dd-liouvillian}
    \mathcal{L}_2[\hat{\rho}] = -\i \sum_{\a, \b} V_\mathbf{ba} \left( \hat{S}^\dagger_\b \hat{S}_\a \hat{\rho} - \hat{S}_\a \hat{\rho} \hat{S}^\dagger_\b \right) + \mathrm{H.c.}
\end{aligned}
\end{equation}
and generic single-spin dynamics governed by $\mathcal{L}_1[\hat{\rho}]$.
Here, $\hat{S}^\dagger_\a = \zeta_\a^\ast \ket{e}_\a \bra{g}_\a$ is the raising operator for the spin at $\a$ with $\zeta_\a$ being an arbitrary phase factor satisfying $|\zeta_\a|^2 = 1$. 
The classical interaction terms yield the effective Hamiltonian under the rotating wave approximation $H_\mathbf{ba} \leftrightarrow \hbar V_\mathbf{ba} \hat{S}^\dagger_\b \hat{S}_\a$, where $\hbar$ is the reduced Planck constant, upon quantization of the dipole moments $\mathbf{d}_\a \leftrightarrow \bra{g}\hat{\mathbf{d}}\ket{e} \zeta_\a \hat{S}^\dagger_\a + \mathrm{H.c.}$, and negating the homogeneous self-interaction energy (Lamb shift) $\Re(H_\mathbf{aa}) \leftrightarrow 0$.
The characteristic energy scale of $\mathcal{L}_2$ is set by the spontaneous decay rate $\Gamma = 2 \Im(V_\mathbf{aa})$.

Eq.~\ref{eqn:dd-liouvillian} has long been known to contain the physics of cooperative decay \cite{Dicke1954} and cooperative Lamb shifts~\cite{Fain1959,Friedberg1973}, with these effects being subsequently observed in a wide variety of physical systems~\cite{Skribanowitz1973,Gross1976,Gross1979,Pavolini1985,Zinovev1983,Varnavskii1984,Devoe1996,Barnes2005,Scheibner2007,Goban2015,McGuyer2015,Houde2017,Garrett1990,Rohlsberger2010,Keaveney2012,vanLoo2013,Meir2014,Ferioli2021}.
More recently, collective electro-magnetic interactions have been considered in the context of ordered arrays of atoms, resulting in a number of novel theoretical predictions~\cite{Chang2004,Zoubi2011,Jenkins2012,Chang2012,Bettles2015,Asenjo-Garcia2017,Henriet2019}, followed by the recent experimental demonstration that a monolayer of atoms on a square lattice can act as an efficient mirror~\cite{Rui2020}.
An emerging paradigm appears to be that the elementary excitations of lattices of quantum mechanical dipoles are hard-core bosons with engineerable dispersion relations~\cite{Asenjo-Garcia2017,Henriet2019,Zhang2019}.
We emphasize that this latter property has no clear classical analogue and is only relevant in the multiple excitation limit $\sum_\a \langle \hat{S}^\dagger_\a\hat{S}_\a \rangle > 1$, distinguishing it from cooperative decay and cooperative Lamb shifts which retain their characteristic behavior in the classical limit. 

The observation of cooperative Lamb shifts in the multiple-excitation limit has only recently been demonstrated~\cite{Glicenstein2020} although the observed shift was shown to disappear before full saturation of the atomic transition due to unknown decoherence mechanisms significantly affecting the dynamics at long interrogation times $\Gamma t \gg 1$.
Three-dimensional optical lattice clocks are natural platforms for studying cooperative electromagnetic interactions, given that all the parameters characterizing $\mathcal{L}_\mathrm{free}$ are systematically characterized, and multi-particle interactions apart from those in $\mathcal{L}_2$ do not significantly affect their evolution.
Despite the presence of a technical dephasing rate $\gamma \gg \Gamma$ due to Raman scattering of optical lattice photons~\cite{Hutson2019}, we rely on the remarkable precision of the atomic clock to divide a Ramsey fringe by more than a part in $10^3$ in order to measure cooperative Lamb shifts with clearly defined excitation fractions $\langle \hat{S}^\dagger_\a \hat{S}_\a\rangle \in [\cos^2(3\pi/8), \cos^2(\pi/8)]$ in the limit $\Gamma t \ll 1$.

\begin{figure}
    \includegraphics[width=3.375in]{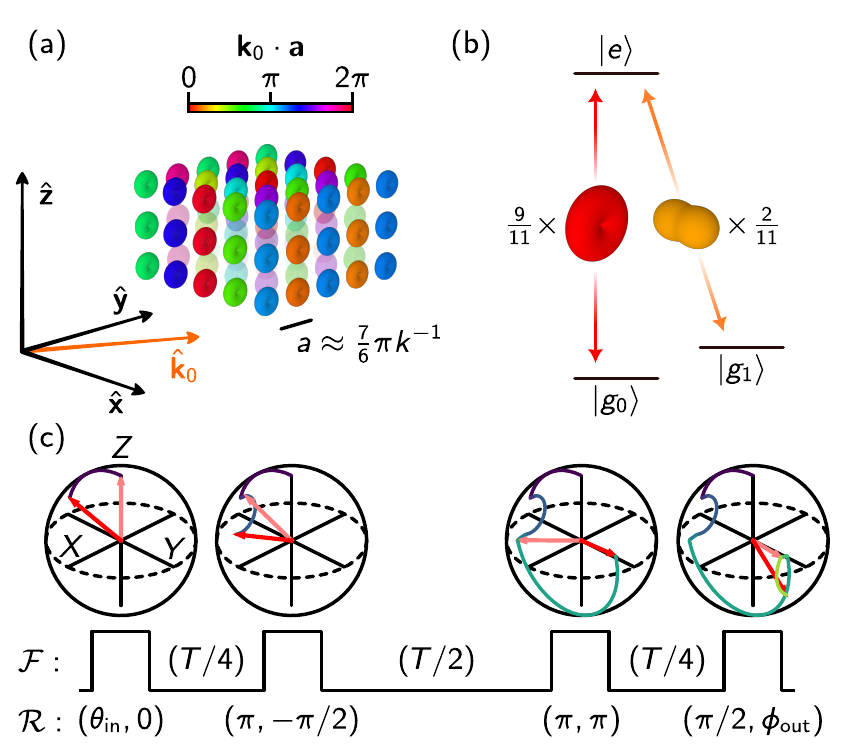}
    \caption{
	    Experimental setup. 
	    (a) Atomic dipoles on a cubic lattice, indexed by their positions $\a$, are excited with a spatially dependent phases $\mathbf{k}_0\cdot\a$. 
		Proximity to the Bragg condition $\mathbf{k}_0 \cdot a \hat{\mathbf{y}} = \pi$ leads to long-range phase ordering along the $\hat{\mathbf{y}}$-axis.
	    (b) Far-field single-atom radiation patterns $\mathcal{V}_q (\hat{\mathbf{r}}_\mathbf{ba})$ for the two spectroscopically resolved $\ket{e} \leftrightarrow \ket{g_q}$ transitions used in this work, as defined in the main text.
	    (c) Resonant laser pulses $\mathcal{R}(\theta, \phi)$ rotate the atomic states by an angle $\theta$ about the $(X_\a \cos\phi + Y_\a \sin\phi)$-axis.
        The various pulses and free-evolution periods $\mathcal{F}(t)$ are chosen such that the output-state projection $\langle \hat{Z}_\a \mathcal{U} \rangle_0$ is proportional only to terms in $\mathcal{L}_\mathrm{free}$ which scale anti-symmetrically with $\cos\theta_\mathrm{in}$, namely those due to resonant dipole-dipole interactions.
    }
    \label{fig:schematic}
\end{figure}

{\em Experimental configuration.---}As previously described in Refs.~\cite{Sonderhouse2020, Milner2023} and schematically represented in Fig.~\subref{fig:schematic}{a}, in a shot-based experiment with a cycle time of $10~\mathrm{s}$, a single-component Fermi-degenerate gas of $N_\mathrm{tot}\approx 9 \times 10^3$ strontium-87 atoms is loaded into the ground band of a cubic optical lattice and initialized into the $\ket{e} = \ket{5s5p~{}^3P_0, F=9/2,m_F = -9/2}$ electronic state.
The optical lattice is formed with a lattice constant of $a_\mathrm{lat} \approx 407~\mathrm{nm}$ by interfering retro-reflected Gaussian laser beams, with $60~\mu\mathrm{m}$ $1/e^{2}$ radii and peak depths of $k_B \times 12~\mu\mathrm{K}$, where $k_B$ is the Boltzmann constant, along each of the $\hat{\mathbf{x}}$-, $\hat{\mathbf{y}}$-, and $\hat{\mathbf{z}}$-axes.
At these depths, tunneling rates are approximately $10~\mathrm{mHz}$ between neighboring sites.
Indexing the lattice sites by their positions $\a = a_\mathrm{lat} (x\hat{\mathbf{x}} + y\hat{\mathbf{y}} + z\hat{\mathbf{z}})$ for integer $\{x, y, z\}$, {\em in-situ} tomographic imaging~\cite{Milner2023} allows for the reconstruction of the site-wise atomic filling fractions $n_\a$, revealing a Fermi-Dirac distribution with a fitted peak density of $n_\mathbf{0} \approx 0.82$, root-mean-square (RMS) radii of $(w_{\hat{\mathbf{x}}}, w_{\hat{\mathbf{y}}}, w_{\hat{\mathbf{z}}}) \approx (3.9~\mu\mathrm{m}, 3.8~\mu\mathrm{m}, 2.1~\mu\mathrm{m})$, and mean entropy per atom of $1.9 k_B$~\cite{Supplement}.

Clock spectroscopy is then performed on the $5s^2~{}^1S_0 \leftrightarrow 5s5p~{}^3P_0$ ``clock'' transition at $\nu = \omega / 2\pi \approx 429~\mathrm{THz}$ using laser-light phase stabilized to a cryogenic-silicon optical cavity~\cite{Matei2017,Oelker2019}.
The probe light propagates with a wave-vector $\mathbf{k}_0 = k \left(\hat{\mathbf{x}}\sin\psi + \hat{\mathbf{y}}\cos\psi\right)$, where $\psi$ is a variable angle of incidence, motivating the choice of local frame $\zeta_\a = e^{ -\i (\omega t - \mathbf{k}_0 \cdot \a) }$.
Resonant pulses with a $2\pi \times 50~\mathrm{Hz}$ Rabi frequency, and variable pulse areas $\theta$ and phase shifts $\phi$, perform global rotations of the atomic state  $\hat{\rho} \rightarrow \mathcal{R}(\theta, \phi) \hat{\rho} = e^{\mathcal{L}_\phi \theta}\hat{\rho}$ with $\mathcal{L}_\phi \hat{\rho} = (-\i / 2) \sum_\a (\hat{S}^\dagger_\a e^{-\i \phi} + \hat{S}_\a e^{\i \phi} ) \hat{\rho} + \mathrm{H.c.}$.
A homogeneous $290~\mu\mathrm{T}$ magnetic field applied along the $\hat{\mathbf{x}}$-axis creates a $540~\mathrm{Hz}$ differential Zeeman splitting between the two available ground states, $\ket{g_q} = \ket{5s^2~{}^1S_0, F=9/2, m_F=-9/2 + q}$ for $q \in \{0,~1\}$, such that their respective resonances with the excited state are spectroscopically resolved.
%As represented in Fig.~\subref{fig:schematic}{b}, each subspace exhibits distinct far-field ($kr_\mathbf{ba} \gg 1$) radiation patterns $I^{(q)}(\hat{\mathbf{r}}_\mathbf{ba}) = (|\mathbf{d}^{(q)}|^2 - |\hat{\mathbf{r}}_\mathbf{ba} \cdot \mathbf{d}^{(q)}|^2) / d^2 \propto (k r_\mathbf{ba}) |V_\mathbf{ba}|$, owing to differences in the magnitudes and orientations of the atomic dipole moments $\mathbf{d}^{(q)} = \bra{e} \hat{\mathbf{d}} \ket{g_q}$: $\mathbf{d}^{(0)} = \sqrt{B^{(0)}} d \hat{\mathbf{x}}$ and $\mathbf{d}^{(1)} = \sqrt{B^{(1)}/2} d \left( \hat{\mathbf{y}} - \i \hat{\mathbf{z}} \right)$~\cite{Cidrim2021}, where $B^{(0)} = 9/11$ ($B^{(1)} = 2/11$) is the spontaneous-decay branching fraction for transition $q=0$ ($q=1$) and $d^2 = \sum_q |\mathbf{d}^{(q)}|^2$ is proportional to the excited state's natural decay rate $\Gamma_0 = k^3 d^2/3\pi\epsilon_0 \hbar = 2\pi \times 1.35(3)~\mathrm{mHz}$~\cite{Muniz2021}.
As represented in Fig.~\subref{fig:schematic}{b}, each subspace exhibits distinct far-field ($kr_\mathbf{ba} \gg 1$) radiation patterns $\mathcal{V}_q (\hat{\mathbf{r}}_\mathbf{ba}) = (k r_\mathbf{ba}) |V_\mathbf{ba}| = 3 \Gamma_\mathrm{nat.} B_q (1 - |\hat{\mathbf{r}}_\mathbf{ba} \cdot \hat{\mathbf{e}}_q|^2) / 4$, owing to differences in the magnitudes and orientations of the atomic dipole moments $\bra{g_q} \hat{\mathbf{d}} \ket{e} = \sqrt{3 \pi \epsilon_0 B_q \Gamma_\mathrm{nat.} / k^3} \mathbf{e}_q$ where $B_0 = 9/11$ ($B_1 = 2/11$) and $\hat{\mathbf{e}}_0 = \hat{\mathbf{x}}$ ($\hat{\mathbf{e}}_1 = [\hat{\mathbf{y}} - \i \hat{\mathbf{z}}] / \sqrt{2}$) are the branching fraction and dipole orientation for transition $q=0$ ($q=1$), respectively~\cite{Cidrim2021}, and $\Gamma_\mathrm{nat.} = 2\pi \times 1.35(3)~\mathrm{mHz}$ is the excited state's natural decay rate ~\cite{Muniz2021}.
%We then explicitly identify the transition-dependent energy scale in Eq.~\ref{eqn:dd-liouvillian} as $\Gamma^{(q)} =  |\mathbf{d}^{(q)}|^2 \Gamma_0$: $\Gamma^{(0)} = 9 \Gamma_0 / 11$ and $\Gamma^{(1)} = 2 \Gamma_0 / 11$.

\begin{figure*}
    \centering
    \includegraphics[width=5.0625in]{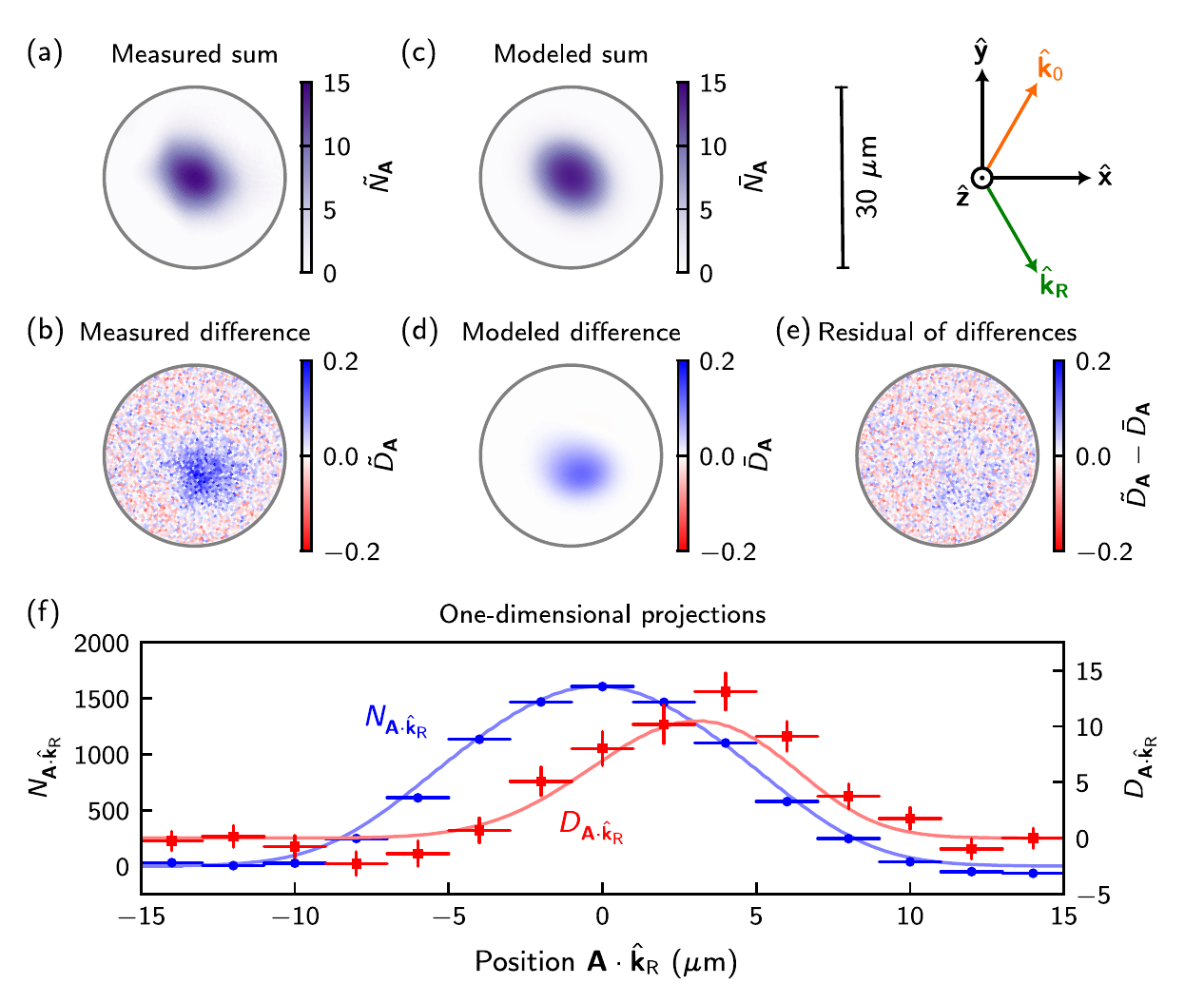}
    \caption{
        Imaging cooperative Lamb shifts at $\psi = \psi_\mathrm{res} = 29.5(5)^\circ$, $\cos\theta_\mathrm{in} = \pm 1/\sqrt{2}$ and $q = 0$.
        Differences in the measured spatial profiles of the sum $\tilde{N}_\mathbf{A}$ (a) and difference $\tilde{D}_\mathbf{A}$ (b) signals indicate the presence of long-range, anisotropic interactions which scale antisymmetrically with $\cos\theta_\mathrm{in}$.
	The maximum of $\tilde{D}_\mathbf{A}$ is spatially offset from the maximum of $\tilde{N}_\mathbf{A}$due to constructive interference of the radiated fields along $\hat{\mathbf{k}}_\mathrm{R} = \mathbf{k}_0 - 2 (\mathbf{k}_0 \cdot \hat{\mathbf{y}})\hat{\mathbf{y}}$.
	The modeled signals $\bar{N}_\mathbf{A}$ (c) and $\bar{D}_\mathbf{A}$ (d) show qualitative agreement with the measured signals upon visual inspection.
	The residuals of the subtraction $\tilde{D}_\mathbf{A} - \bar{D}_\mathbf{A}$ are shown in (e).
	The modeled signals are obtained by fitting $\tilde{N}_\mathbf{A}$ to a Fermi-Dirac distribution and contain no other free parameters.
        (f) One-dimensional projections $I_{\mathbf{A} \cdot \hat{\mathbf{k}}_\mathrm{R}}$ of the above signals are obtained by projecting the images $I_\mathbf{A}$ in panels (a-d) onto $\mathbf{k}_\mathrm{R}$, {\em i.e.} $I_{\mathbf{A} \cdot \hat{\mathbf{k}}_\mathrm{R}} = \sum_{\mathbf{A}' \parallel (\mathbf{A} \cdot \hat{\mathbf{k}}_\mathrm{R}) \hat{\mathbf{k}}_\mathrm{R} } I_{\mathbf{A}'}$.
        The data points display the measured signals with $\tilde{N}_\mathbf{A}$ as blue circles and $\tilde{D}_\mathbf{A}$ as red squares. 
	Vertical error bars represent $1\sigma$ standard errors and horizontal bars show the $2~\mu\mathrm{m}$ bin-width of the projections onto $\hat{\mathbf{k}}_\mathrm{R}$.
        The red (blue) solid line shows the modeled signal $\bar{N}_{\mathbf{A}\cdot \hat{\mathbf{k}}_\mathrm{R}}$ ($\bar{D}_{\mathbf{A}\cdot \hat{\mathbf{k}}_\mathrm{R}}$).
    }
    \label{fig:images}
\end{figure*}

Denoting the free-evolution generated by $\mathcal{L}_\mathrm{free}$ as $\mathcal{F}(t) = e^{\mathcal{L}_\mathrm{free} t}$, the spectroscopic sequence can be represented as 
\begin{equation}
\begin{aligned}
    \mathcal{U} = &\mathcal{R}(\pi/2, \phi_\mathrm{out}) \mathcal{F}(T/4) \mathcal{R}(\pi, \pi) \mathcal{F}(T/2) \\
    & \times \mathcal{R}(\pi, -\pi/2) \mathcal{F}(T/4) \mathcal{R}(\theta_\mathrm{in}, 0)
 \end{aligned}
\end{equation}
where $\theta_\mathrm{in}$ is the variable initial pulse area, $T = 2~\mathrm{s}$ is the total free-evolution period, and $\phi_\mathrm{out}$ is the variable final pulse phase.
Fig.~\subref{fig:schematic}{c} depicts the time evolution of the atomic state throughout $\mathcal{U}$, as represented on the Bloch sphere with vector components $\hat{X}_\a = \hat{S}^\dagger_\a + \hat{S}_\a$, $\hat{Y}_\a = -\i ( \hat{S}^\dagger_\a - \hat{S}_\a )$, and $\hat{Z}_\a = \hat{S}^\dagger_\a \hat{S}_\a - \hat{S}_\a \hat{S}^\dagger_\a $.
Starting with the initial conditions $\langle \hat{S}^\dagger_\a \hat{S}_\a \rangle_0 = n_\a$ and $\langle \hat{S}_\a \hat{S}^\dagger_\a \rangle_0 = 0$, the first pulse $\mathcal{R}(\theta_\mathrm{in}, 0)$ rotates the population imbalance of the atomic state $\langle \hat{Z}_\a \rangle \rightarrow n_\a \cos\theta_\mathrm{in}$.
The elastic contribution of $\mathcal{L}_2$ to the subsequent free-evolution can be intuited, for short times $\Gamma t \ll 1$, as an Ising-type interaction which rotates each atom about the $Z_\a$-axis at a rate $2 \sum_\b \Re(V_\mathbf{ba}) \langle \hat{S}^\dagger_\b \hat{S}_\a \rangle$.

The two ``spin-echo'' pulses  $\mathcal{R}(\pi, \phi)$ preserve the coherent dynamics generated by $\mathcal{L}_2$ while suppressing the various single-particle dephasing mechanisms contained in $\mathcal{L}_1[\hat{\rho}]  = -\i \sum_\a (\Delta\omega_a - \i \gamma / 2)(\hat{S}^\dagger_\a \hat{S}_\a \hat{\rho} - \hat{S}_\a \hat{\rho} \hat{S}^\dagger_\a ) + \mathrm{H.c.}$, where $\Delta \omega_\a$ is the relative detuning of atom $\a$ with respect to the probe laser, and $\gamma/2 \approx (9.3~\mathrm{s})^{-1}$ is the homogeneous dephasing rate due to Raman scattering of optical lattice photons~\cite{Hutson2019}. 
The dominant contribution to $\Delta\omega_\a$ arises from frequency drifts of the probe laser on the order of $1~\mathrm{Hz}$ between daily measurements of the transition resonance frequencies.
Differential ac Stark shifts varying with the local lattice intensity also contribute to $\Delta\omega_\a$, yet are limited to the sub-$10~\mathrm{mHz}$ level by optimizing the optical frequency of each trapping beam~\cite{Campbell2017}.
These detunings do not directly affect the final state since the spin-echo pulses anti-commute with time-evolution under $\mathcal{L}_1$ in the limit $\gamma \rightarrow 0$, whereas the spin-echo pulses approximately commute with evolution under $\mathcal{L}_2$ for $\Gamma t\ll 1$.
Finite $\gamma$ leads to a decay in both the single atom coherences $\langle \hat{S}^\dagger_\a \rangle \propto e^{-\gamma t / 2}$, and excited state populations $\langle \hat{S}^\dagger_\a \hat{S}_\a \rangle \propto e^{-\gamma t}$.
For an increasing number of spin-echo pulses, the time-averaged longitudinal decay asymptotically approaches $\langle \hat{Z}_\a \rangle \propto e^{-\gamma t / 2}$.

The final $\mathcal{R}(\pi/2, \phi_\mathrm{out})$-pulse maps the interaction-induced phase shifts of the coherences onto the difference in electronic populations.
A diffraction limited imaging system with a $1.3~\mu\mathrm{m}$ resolution then records the column-integrated populations of the ground $\tilde{N}^{g}_\mathbf{A}$ and excited $\tilde{N}^{e}_\mathbf{A}$ states via absorption imaging onto a sCMOS camera where $\mathbf{A} = a_\mathrm{px} ( x \hat{\mathbf{x}}_\mathrm{px} + y \hat{\mathbf{y}}_\mathrm{px} )$ indexes the sensor pixels with grid-axes $\hat{\mathbf{x}}_\mathrm{px}$ and $\hat{\mathbf{y}}_\mathrm{px}$ rotated with respect to the lattice axes by $30^\circ$ about the $\hat{\mathbf{z}}$-axis, and $a_\mathrm{px} \approx 410~\mathrm{nm}$ is the effective pixel size~\cite{Marti2018}.

Approximately $2 \times 10^4$ experimental shots were recorded over a period of 2 weeks while independently modulating the four parameters $\sin\phi_\mathrm{out} \in \{-1,+1\}$, $\cos\theta_\mathrm{in} \in \{-4, -3, \ldots, 4 \} / 4\sqrt{2}$, $q \in \{0,~1\}$, and $\psi \in \{0^\circ, \psi_\mathrm{res} = 29.5(5)^\circ\}$.
Fig.~\subref{fig:images}{a-b} displays the sum $\tilde{N}_\mathbf{A} = (\tilde{N}_\mathbf{A}^e + \tilde{N}_\mathbf{A}^g)$ and difference $\tilde{D}_\mathbf{A} = (\tilde{N}_\mathbf{A}^e - \tilde{N}_\mathbf{A}^g) / C$ signals, with $C = \sin(\theta_\mathrm{in}) \sin(\phi_\mathrm{out}) e^{-\gamma T / 2} \approx -\partial_{\phi_\mathrm{out}} \tilde{D}_\mathbf{A} / \tilde{N}_\mathbf{A}$ being the interferometric sensitivity, averaged over the subset of data with maximal interaction strengths: $\psi = \psi_\mathrm{res} $, $\cos\theta_\mathrm{in} = \pm 1/\sqrt{2}$ and $q = 0 $.

\begin{figure}
    \includegraphics[width=3.375in]{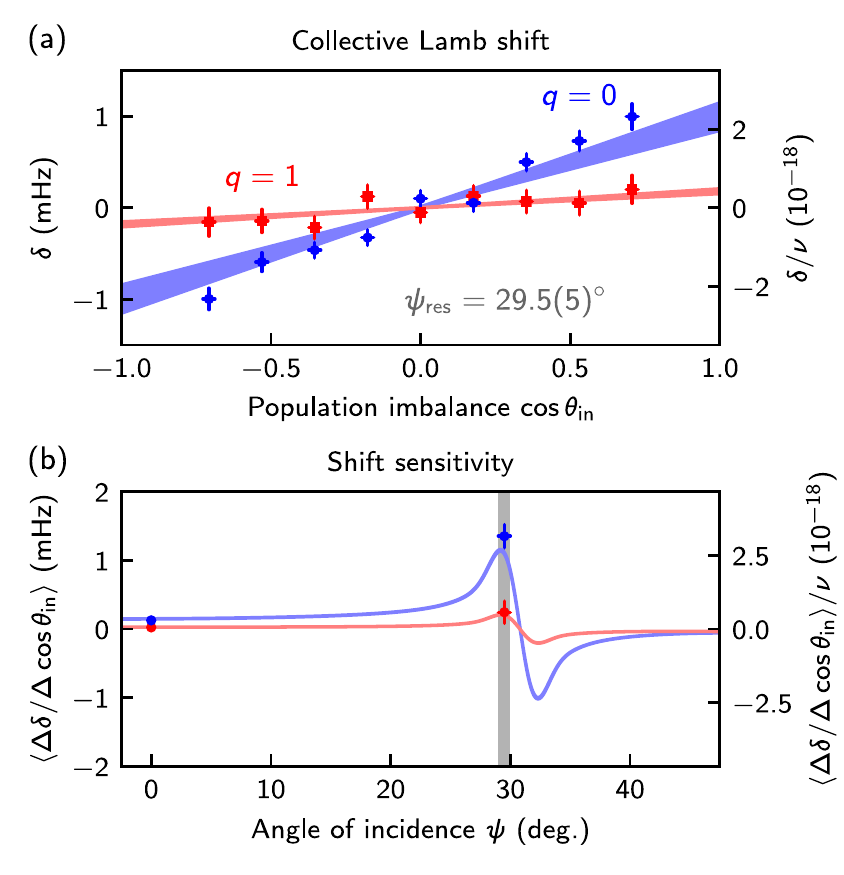}
    \caption{
	    Controlling cooperative Lamb shifts. 
	    (a) Scaling of the ensemble-averaged shift $\delta$ versus the initial spin projection $\cos\theta_\mathrm{in}$ for $\psi = \psi_\mathrm{res}$ and both $q=0$ (blue) and $q=1$ (red).
	    Data points show the measured shifts $\tilde{\delta}$ with vertical error bars representing $1\sigma$ standard errors and horizontal error bars representing $2\%$ observed fluctuations in the pulse areas $\theta$.
	    Shaded regions show the modeled shifts $\bar{\delta}$, propagating the experimental uncertainty in $\psi_\mathrm{res}$.
	    (b) Angle of incidence dependence of the shift sensitivity to changes in the initial spin projection. 
            The vertical gray bar represents the angle of incidence used in (a).
    }
    \label{fig:shifts}
\end{figure}

{\em Analysis.---}We compare these and other experimentally derived quantities, denoted by symbols covered with a tilde $\tilde{\cdot}$, to their modeled equivalents, denoted by symbols covered with a bar $\bar{\cdot}$, obtained by substituting $\tilde{N}^g_\mathbf{A} \rightarrow \bar{N}^g_\mathbf{A}  = \sum_{\a\parallel\mathbf{A}} (n_\a - \langle \hat{Z}_\a\mathcal{U}\rangle_0) / 2$ and $\tilde{N}^e_\mathbf{A} \rightarrow \bar{N}^e_\mathbf{A} = \sum_{\a\parallel\mathbf{A}} (n_\a + \langle \hat{Z}_\a\mathcal{U}\rangle_0) / 2$ where $\a \parallel \mathbf{A}$ denotes the set of all atoms whose image is projected onto the pixel $\mathbf{A}$, and the expectation value of $\hat{Z}_\a$ with respect to the final state is computed as
\begin{equation}
\begin{aligned}
     \langle \hat{Z}_\a \mathcal{U} \rangle_0 &= n_\a C \left[ J_\a \cos\Phi + \left( 1 - K_\a \right) \sin\Phi \right]\\
J_\a &= \cos(\theta_\mathrm{in}) T \sum_{\b \neq \a} n_\b \Re\left(V_\mathbf{ba}\right) + \mathcal{O}(\Gamma \gamma T^2)\\
K_\a &= \frac{\Gamma T}{2} + \mathcal{O}(\Gamma \gamma T^2)
\end{aligned}
\end{equation}
where $J_\a$ ($K_\a$) is the leading order, in $\Gamma T$, phase-shift (decoherence) of atom $\a$ due to resonant dipole-dipole interactions with all other atoms and whose full time-dependence in terms of the quantity $\gamma T$ is given in Ref.~\cite{Supplement}.
The parameter $\Phi = \Delta\phi(T) - 2 \Delta\phi(3 T / 4) + 2 \Delta\phi(T/4) - \Delta\phi(0)$ results from propogating noise-induced deviations in the time-dependent probe-laser phase $\phi \rightarrow \phi + \Delta\phi(t)$ evaluated at each of the four rotation pulses.
On timescales comparable to $T$, the probe laser exhibits white frequency noise, where the RMS difference in phases over a time interval $\Delta t$ is approximately $\sqrt{\langle \Delta\phi^2(\Delta t)\rangle} / \Delta t \approx 90~\mathrm{mrad.}~\mathrm{s}^{-1}$~\cite{Matei2017}, contributing a zero-mean, stochastic signal on the order of $\Delta\tilde{D}_\mathbf{A} \approx \sqrt{3 \langle \Delta \phi^2(T) \rangle / 8} \tilde{N}_\mathbf{A}$ to individual measurements of $\tilde{D}_\mathbf{A}$~\cite{Supplement}.

Owing to differences in the spatial profiles of $\bar{N}_\mathbf{A}\propto n_\a$ and $\bar{D}_\mathbf{A}\propto n_\a J_\a$, we are able to remove population differences due to fluctuations in the probe-laser phase by applying corrections $\tilde{D}_\mathbf{A} \rightarrow \tilde{D}_\mathbf{A} - \mathcal{P}^\mathrm{opt}_N\tilde{N}_\mathbf{A}$ to the presented data, where the coefficients $\mathcal{P}^\mathrm{opt}_N$ are obtained from least squares fits minimizing the quantity $\sum_\mathbf{A} (\mathcal{P}_N \bar{N}_\mathbf{A} + \mathcal{P}_D\bar{D}_\mathbf{A} - \tilde{D}_\mathbf{A})^2 / \mathrm{Var}(\tilde{D}_\mathbf{A})$ over the parameters $\mathcal{P}_N$ and $\mathcal{P}_D$. 

Fig.~\subref{fig:images}{c-f} shows the spatial profile and absolute scale of the modeled quantities $\bar{N}_\mathbf{A}$ and $\bar{D}_\mathbf{A}$, averaged over $\Phi$, to be in good agreement with the measurement.
The ensemble-averaged cooperative Lamb shifts $\tilde{\delta} = \sum_\mathbf{A} \tilde{D}_\mathbf{A} / 2 \pi T \sum_\mathbf{A} \tilde{N}_\mathbf{A}$ are plotted against $\cos\theta_\mathrm{in}$ for $\psi = \psi_\mathrm{res}$ in Fig.~\subref{fig:shifts}{a}.
A $\chi^2$ analysis comparing the shifts evaluated at each set of $(\cos \theta_\mathrm{in},q,\psi)$ to the model gives $\chi^2 / (20~\mathrm{d.o.f}) \approx 1.1$.

\begin{figure}
    \includegraphics[width=3.375in]{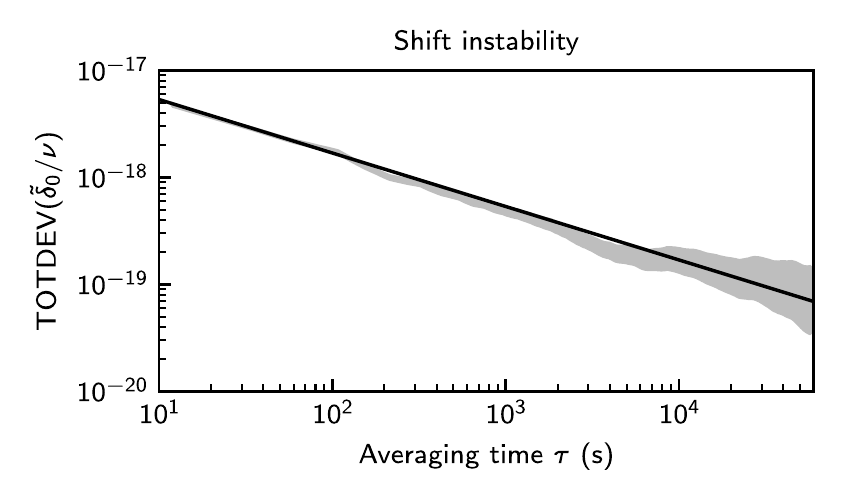}
    \caption{
        Instability of clock shift evaluation.
        The shaded gray region shows the $1\sigma$ confidence interval of the total deviation~\cite{Howe2000} of the clock shift $\tilde{\delta}_0 / \nu$ as a function of averaging time $\tau$.
        The black line shows a fit to the data for $\tau < 100~\mathrm{s}$, assuming a white noise floor.
    }
    \label{fig:instability}
\end{figure}

While $V_\mathbf{ba}$ asymptotically decays with increasing separation as $1 / kr_\mathbf{ba}$, the contained phase-factors $e^{-\i (kr_\mathbf{ba} + \mathbf{k}_0\cdot\mathbf{r}_\mathbf{ba})}$ average to zero for incommensurate $ka_\mathrm{lat} \approx 7\pi/6$ and $\psi = 0^\circ$, resulting in effectively nearest neighbor interactions scaling with the local filling fractions, {\em i.e.} $J_\a \propto n_\a$.
However, the sub-wavelength lattice spacing $ka_\mathrm{lat} < 2\pi$ guarantees the unique existence of the Bragg resonance at $\psi = \arccos (\pi / ka_\mathrm{lat}) \approx 30.8^\circ$ satisfying $\mathbf{k}_0 \cdot a \hat{\mathbf{y}} = \pi$ such that the radiated fields add constructively along $\mathbf{k}_\mathrm{R} = \mathbf{k}_0 - 2 (\mathbf{k}_0 \cdot \hat{\mathbf{y}})\hat{\mathbf{y}}$.
Numerically, we find that the ensemble-averaged interaction strengths are maximized, and scale with the system size as $N_\mathrm{tot}^{1/3}$, at angular detunings from exact Bragg resonance set by the diffraction limit $\pi / 2 k ( w^2_{\hat{\mathbf{x}}} + w^2_{\hat{\mathbf{y}}} )^{1/2} \approx 1.9^\circ$~\cite{Chang2004,Manassah2010,Supplement}.
The shift's sensitivity to changes in the initial tipping angle $\langle \Delta\tilde{\delta} / \Delta\cos\theta_\mathrm{in} \rangle$, averaged over all observed $\theta_\mathrm{in}$, is plotted against $\psi$ in Fig.~\subref{fig:shifts}{b}. 

Extrapolating to clock operation at $\cos\theta_\mathrm{in} = 0$ with $q=0$, and $\psi = \psi_\mathrm{res}$, we evaluate the systematic shift due to resonant dipole-dipole interactions to be $\tilde{\delta}_0 / \nu = \sum_{\theta_\mathrm{in}} \tilde{\delta} / \sum_{\theta_\mathrm{in}} \nu = -1.5(7)\times 10^{-19}$ in fractional frequency units, demonstrating that even for the most sensitive geometry considered in this work, systematic effects can be made negligible relative to the lowest reported total systematic uncertainties for optical atomic clocks~\cite{McGrew2018,Brewer2019,Bothwell2019}.
Fig.~\ref{fig:instability} displays the fractional frequency instability of the $\tilde{\delta}_0 / \nu$ evaluation.
A fit to the data reveals a $1.7 \times 10^{-17} / \sqrt{\mathrm{Hz}}$ short-term white-noise floor.

{\em Conclusion.---}We have performed measurements of, and successfully modeled cooperative Lamb shifts in a three-dimensional optical lattice clock. 
Control over the spatial orientations of the probe light, and excited dipole moments allows for a dramatic modification of the magnitude of these effects---from levels relevant to state-of-the-art atomic clocks to more than an order of magnitude below. 
Technical dephasing due to Raman scattering of optical lattice photons prevented the study of dynamics beyond $\Gamma t \ll 1$.
It is interesting to consider future work where collective interactions can be made significantly stronger than technical dephasing rates by either probing transitions with stronger intrinsic dipole moments or longer transition wavelengths, or by optically dressing the excited state used in this work~\cite{Santra2005}.
Under such conditions, the dynamics of collective light-matter interactions are expected to lead to spin-squeezing~\cite{Chunlei2019} and other exotic states of quantum matter~\cite{Henriet2019}.

{\em Acknowledgements---}We thank D. Kedar for technical assistance, and D. E. Chang, H. Ritsch and M. D. Lukin useful discussions.  
We also thank  S. L. Campbell, N. Darkwah Oppong, A. M. Rey and D. Wellnitz for careful reading of the manuscript and for providing insightful comments. 
{\bf Funding:} Funding for this work is provided by NSF QLCI OMA-2016244, DOE Center of Quantum System Accelerator, V. Bush Fellowship, NIST, and NSF Phys-1734006.
{\bf Author Contributions:} All authors contributed to carrying out the experiments, interpreting the results, and writing the manuscript.
{\bf Competing interests:} The authors declare no competing interests.
{\bf Data and materials availability:} Data from the main text and supplementary materials are available from the corresponding authors upon reasonable request.

%\bibliographystyle{apsrev4-2}
%\bibliography{references}


\clearpage 

\onecolumngrid
\section{Supplemental materials}

\subsection{Atomic density distribution}
The center of mass of the atomic cloud is trapped by Gaussian laser beams with intensity profiles $I^{\hat{\mathbf{n}}}_\mathbf{a} = I^{\hat{\mathbf{n}}}_0 e^{-2 |\mathbf{a}_{\hat{\mathbf{n}}}|^2 / W_{\hat{\mathbf{n}}}^2}$ where the unit vectors of $\mathbf{n} \in \{\mathbf{x}, ~\mathbf{y}, ~\mathbf{z}\}$ index the three lattice axes, $\mathbf{a}_{\hat{\mathbf{n}}} = \mathbf{a} - (\hat{\mathbf{n}}\cdot\mathbf{a})\hat{\mathbf{n}}$ is the transverse position with respect to $\hat{\mathbf{n}}$-axis, $W_{\hat{\mathbf{n}}}$ is the $\hat{\mathbf{n}}$-th beam's $e^{-2}$ radius, and $I^{\mathbf{n}}_0$ is the $\hat{\mathbf{n}}$-th beams peak intensity.
Summing over all beams and multiplying by the ac polarizability $\alpha$ gives the potential energy
\begin{equation}
\begin{aligned}
    U_\mathbf{a} = -\alpha \sum_{\hat{\mathbf{n}}}I_\mathbf{a}^{\hat{\mathbf{n}}}
        \approx -\alpha \sum_{\hat{\mathbf{n}}}I^{\hat{\mathbf{n}}}_0  \left(1 - 2\frac{|\mathbf{a}_{\hat{\mathbf{n}}}|^2}{W_{\hat{\mathbf{n}}}^2}\right)
\end{aligned}
\end{equation}
where we have approximated $|\mathbf{a}_{\hat{\mathbf{n}}}| \ll W_{\hat{\mathbf{n}}}$.

This motivates us defining the dimensionless potential $u_\mathbf{a} = \mathbf{a} \cdot \mathbf{u} \cdot \mathbf{a}$ with
\begin{equation}
    \mathbf{u} = \frac{1}{2a_\mathrm{lat}^2} \left( \sigma^{-2}_{\hat{\mathbf{x}}} ~ \hat{\mathbf{x}} \hat{\mathbf{x}} + \sigma^{-2}_{\hat{\mathbf{y}}} ~ \hat{\mathbf{y}} \hat{\mathbf{y}} + \sigma^{-2}_{\hat{\mathbf{z}}} ~ \hat{\mathbf{z}} \hat{\mathbf{z}} \right)
\end{equation}
for the purpose of fitting density distributions $n_\mathbf{a}$. 
The dimensionless fit parameters $\sigma_{\hat{\mathbf{n}}}$ characterize the strength of the confining potential along the $\hat{\mathbf{n}}$-th axis.
For a grand canonical ensemble under the local-density-approximation we define the dimensionless local chemical potential $\mu_\mathbf{a} = \mu_\mathbf{0} - u_\mathbf{a}$ where the peak chemical potential $\mu_\mathbf{0}$ is an additional fit parameter.
The corresponding partition function for identical fermions restricted to the ground band of the lattice is given by $Z_\mathbf{a} = 1 + e^{\mu_\mathbf{a}}$.
The local density $n_\mathbf{a}$ is then
\begin{equation}
    n_\mathbf{a} = Z_\mathbf{a}^{-1} e^{\mu_\mathbf{a}} = \frac{e^{\mu_\mathbf{a}}}{1 + e^{\mu_\mathbf{a}}} ~ .
\end{equation}
Summing along the imaging axis $\hat{\mathbf{m}}$ gives the column density fitting function
\begin{equation}
    N_{\mathbf{a}_{\hat{\mathbf{m}}}} = \sum_{\mathbf{a}' \parallel \mathbf{a}_{\hat{\mathbf{m}}}} n_{\mathbf{a}'} ~ .
\end{equation}

Fitting absorption images to $N_{\mathbf{a}_{\hat{\mathbf{m}}}}$ with four free parameters $( \mu_\mathbf{0}, \sigma_{\hat{\mathbf{x}}}, \sigma_{\hat{\mathbf{y}}}, \sigma_{\hat{\mathbf{z}}} )$ exhibits large covariance between $\mu_\mathbf{0}$ and $w_{\hat{\mathbf{n}}}$ for $\mu_\mathbf{0} \ll 1$.
This ambiguity can be fixed by imaging along an orthogonal axis.
Images taken along $\hat{\mathbf{m}} = \hat{\mathbf{z}}$ reveal $\sigma_{\hat{\mathbf{x}}} \approx \sigma_{\hat{\mathbf{y}}}$ such that the aspect ratio $\beta = \sqrt{\sigma_{\hat{\mathbf{x}}} \sigma_{\hat{\mathbf{y}}}} / \sigma_{\hat{\mathbf{z}}} \approx 1.79$ can be constrained by fitting absorption images taken along $\hat{\mathbf{m}} = \cos(\pi/6) \hat{\mathbf{x}} + \sin(\pi/6) \hat{\mathbf{y}}$.
Under this constraint, the fit to $N_{\mathbf{a}_{\hat{\mathbf{z}}}}$, yields $\mu_0 \approx 1.44$ , $\sigma_{\hat{\mathbf{x}}} \approx 8.8$, and $\sigma_{\hat{\mathbf{y}}} \approx 7.2$.

With our model for $n_\mathbf{a}$ determined, various thermodynamic quantities can be numerically extracted.
The peak density is
\begin{equation}
    n_\mathbf{0} = \frac{e^{\mu_0}}{1 + e^{\mu_0}} \approx 0.81~.
\end{equation}
The total atom number is
\begin{equation}
    N = \sum_\mathbf{a} n_\mathbf{a} \approx 9 \times 10^{3}~,
\end{equation}
in agreement with summed counts from both {\em in-situ} and time-of-flight absorption measurements.
The RMS width along the $\hat{\mathbf{n}}$-th axis is given by
\begin{equation}
    w_{\hat{\mathbf{n}}}^2 = \sum_\mathbf{a} \left(\mathbf{a} \cdot \hat{\mathbf{n}}\right)^2 n_\mathbf{a} \rightarrow (w_{\hat{\mathbf{x}}}, w_{\hat{\mathbf{y}}}, w_{\hat{\mathbf{z}}}) = (3.9~\mu\mathrm{m}, 3.8~\mu\mathrm{m}, 2.1~\mu\mathrm{m})~.
\end{equation}
The entropy per particle at the $\mathbf{a}$-th lattice site is
\begin{equation}
    s_\mathbf{a} = k_\mathrm{B} \left[n_\mathbf{a} \log n_\mathbf{a} + (1 - n_\mathbf{a}) \log(1 - n_\mathbf{a})\right] / n_\mathbf{a}.
\end{equation}
for the central lattice site, $s_0 \approx 0.6 k_\mathrm{B}$.
Summing over all sites, the mean entropy per particle is
\begin{equation}
    S = \sum_\mathbf{a} n_\mathbf{a} s_\mathbf{a} / N \approx 1.9 k_\mathrm{B} ~ .
\end{equation}
The (dimensionless) average potential energy is
\begin{equation}
    \epsilon = \sum_\mathbf{a} n_\mathbf{a} u_\mathbf{a}\approx 1.4 ~ .
\end{equation}
The (dimensionless) Fermi energy $\epsilon_F$ is implicitly defined via 
\begin{equation}
    N = \sum_{u_\mathbf{a} \leq \epsilon_F} 1 \rightarrow \epsilon_F \approx 2.0 ~ ,
\end{equation}
yielding a degeneracy parameter 
\begin{equation}
    \epsilon/\epsilon_F \approx 0.7 ~ .
\end{equation}

\subsection{Site-wise interferometer signal}
The pulses $\mathcal{R}(\theta, \phi)$ evolve the single-particle expectation values according to
\begin{equation}
\begin{aligned}
    \partial_\theta \langle \hat{S}^\dagger_\mathbf{a} \rangle &= \partial_\theta \langle \hat{S}_\mathbf{a} \rangle^\ast = \Tr\left(\hat{S}^\dagger_\mathbf{a} \mathcal{L}_\mathrm{rot}[\hat{\rho}] \right) = - \frac{\i}{2} e^{\i \phi} \left( \langle \hat{S}^\dagger_\mathbf{a} \hat{S}_\mathbf{a} \rangle  - \langle \hat{S}_\mathbf{a} \hat{S}^\dagger_\mathbf{a} \rangle \right)\\
    \partial_\theta \langle \hat{S}^\dagger_\mathbf{a} \hat{S}_\mathbf{a} \rangle &= -\partial_\theta \langle \hat{S}_\mathbf{a} \hat{S}^\dagger_\mathbf{a} \rangle = \Tr\left(\hat{S}^\dagger_\mathbf{a} \hat{S}_\mathbf{a} \mathcal{L}_\mathrm{rot}[\hat{\rho}] \right) = - \frac{\i}{2} \left( e^{-\i \phi} \langle \hat{S}^\dagger_\mathbf{a} \rangle - e^{\i \phi} \langle \hat{S}_\mathbf{a} \rangle \right ) ~ .
\end{aligned}
\end{equation}
These coupled differential equations can be solved to obtain the single-particle matrix elements of $\mathcal{R}$,
\begin{equation}
\begin{aligned}
    \langle \hat{S}^\dagger_\mathbf{a} \mathcal{R}(\theta, \phi) \rangle &= \cos^2(\theta / 2 ) \langle \hat{S}^\dagger_\mathbf{a} \rangle + e^{2 \i \phi}\sin^2(\theta / 2) \langle \hat{S}_\mathbf{a} \rangle - \frac{\i}{2}e^{\i \phi} \sin(\theta) \left( \langle \hat{S}^\dagger_\mathbf{a} \hat{S}_\mathbf{a} \rangle - \langle \hat{S}_\mathbf{a} \hat{S}^\dagger_\mathbf{a} \rangle \right) \\
    \langle \hat{S}^\dagger_\mathbf{a} \hat{S}_\mathbf{a} \mathcal{R}(\theta, \phi) \rangle &= \cos^2(\theta/2) \langle \hat{S}^\dagger_\mathbf{a}\hat{S}_\mathbf{a} \rangle + \sin^2(\theta/2) \langle \hat{S}_\mathbf{a}\hat{S}^\dagger_\mathbf{a} \rangle - \frac{\i}{2} \sin(\theta) \left( e^{-\i \phi} \langle \hat{S}^\dagger_\mathbf{a} \rangle - e^{\i \phi} \langle \hat{S}_\mathbf{a} \rangle \right)\\
    \langle \hat{S}_\mathbf{a} \hat{S}^\dagger_\mathbf{a} \mathcal{R}(\theta, \phi) \rangle &= \cos^2(\theta/2) \langle \hat{S}_\mathbf{a}\hat{S}^\dagger_\mathbf{a} \rangle + \sin^2(\theta/2) \langle \hat{S}^\dagger_\mathbf{a}\hat{S}_\mathbf{a} \rangle + \frac{\i}{2} \sin(\theta) \left( e^{-\i \phi} \langle \hat{S}^\dagger_\mathbf{a} \rangle - e^{\i \phi} \langle \hat{S}_\mathbf{a} \rangle \right)~.
\end{aligned}
\end{equation}
Since $\mathcal{L}_\mathrm{rot}$ is a sum over single particle operators, multi-particle matrix elements of $\mathcal{R}$ can be obtained directly from the above elements, {\em e.g.} $\partial\langle \hat{{O}}_\mathbf{a} \hat{{O}}_\mathbf{b} \mathcal{R}(\theta, \phi)\rangle / \partial \langle \hat{{P}}_\mathbf{a} \hat{{P}}_\mathbf{b} \rangle  = \partial^2 \langle \hat{{O}}_\mathbf{a} \mathcal{R}(\theta, \phi)\rangle \langle \hat{{O}}_\mathbf{b} \mathcal{R}(\theta, \phi)\rangle / \partial \langle \hat{{P}}_\mathbf{a} \rangle \partial \langle \hat{{P}}_\mathbf{b} \rangle$.
In particular, we will make use of 
\begin{equation}
\begin{aligned}
    \langle \hat{S}^\dagger_\mathbf{a} \mathcal{R}(\theta, 0)\rangle_0 &= - \frac{\i}{2} \sin(\theta) n_\mathbf{a}\\
    \langle \hat{S}^\dagger_\mathbf{a} \hat{S}_\mathbf{a} \hat{S}^\dagger_\mathbf{b} \mathcal{R}(\theta, 0)\rangle_0 &= -\frac{\i}{2} \sin(\theta) \cos^2(\theta/2) n_\mathbf{a} n_\mathbf{b} \quad \mathrm{for }~ \mathbf{a} \neq \mathbf{b} \\ 
    \langle \hat{S}_\mathbf{a} \hat{S}^\dagger_\mathbf{a} \hat{S}^\dagger_\mathbf{b} \mathcal{R}(\theta, 0)\rangle_0 &= - \frac{\i}{2} \sin(\theta) \sin^2(\theta/2) n_\mathbf{a} n_\mathbf{b} \quad \mathrm{for }~ \mathbf{a} \neq \mathbf{b} \\
    \langle \hat{S}^\dagger_\mathbf{a} \mathcal{R}(\pi, \phi)\rangle &= e^{2\i \phi} \langle \hat{S}_\mathbf{a} \rangle\\ 
    \langle \hat{S}^\dagger_\mathbf{a} \hat{S}_\mathbf{a} \hat{S}^\dagger_\mathbf{b} \mathcal{R}(\pi, \phi)\rangle &= e^{2\i \phi} \langle \hat{S}_\mathbf{a} \hat{S}^\dagger_\mathbf{a} \hat{S}_\mathbf{b} \rangle ~ \mathrm{for } \quad \mathbf{a} \neq \mathbf{b} \\ 
    \langle \hat{S}_\mathbf{a} \hat{S}^\dagger_\mathbf{a} \hat{S}^\dagger_\mathbf{b} \mathcal{R}(\pi, \phi)\rangle &= e^{2 \i \phi} \langle \hat{S}^\dagger_\mathbf{a} \hat{S}_\mathbf{a} \hat{S}_\mathbf{b} \rangle ~ \mathrm{for } \quad \mathbf{a} \neq \mathbf{b} \\
\langle \hat{Z}_\mathbf{a} \mathcal{R}(\pi/2, \phi)\rangle &= - \i \left[ e^{-\i \phi} \langle \hat{S}^\dagger_\mathbf{a} \rangle  - e^{\i \phi} \langle \hat{S}_\mathbf{a} \rangle \right]~.
\end{aligned}
\end{equation}

The free-evolution of the coherence $\langle \hat{S}^\dagger_\mathbf{a} \rangle$ is given by
\begin{equation}
\begin{aligned}
    \partial_t \langle \hat{S}^\dagger_\mathbf{a} \rangle &= \Tr(\hat{S}^\dagger_\mathbf{a} \mathcal{L}_\mathrm{free}[\hat{\rho}] ) = \i \sum_i \sum_{\mathbf{b}} h_\mathbf{ba}^{(i)} \left( \langle \hat{S}^\dagger_\mathbf{a} \hat{S}_\mathbf{a} \hat{S}^\dagger_\mathbf{b} \rangle  - \langle \hat{S}_\mathbf{a} \hat{S}^\dagger_\mathbf{a} \hat{S}^\dagger_\mathbf{b} \rangle \right)
\end{aligned}
\end{equation}
where 
\begin{equation}
\begin{aligned}
    h^{(1)}_\mathbf{ba} &= (\Delta \omega_\mathbf{a} - \i \gamma / 2) \delta_{\mathbf{a},\mathbf{b}}\\
    h^{(2)}_\mathbf{ba} &= V_\mathbf{ba}\\
        &= - \frac{3\Gamma_\mathrm{nat.} B_q}{4} \left\{  \left[ 1 - |\hat{\mathbf{r}}_\mathbf{ba} \cdot \hat{\mathbf{e}}_q|^2 \right] \frac{1}{kr_\mathbf{ba}} + \left[ 3 |\hat{\mathbf{r}}_\mathbf{ba} \cdot \hat{\mathbf{e}}_q|^2 - 1 \right] \left[\frac{1}{(kr_\mathbf{ba})^3} + \frac{\i}{(kr_\mathbf{ba})^2}\right] \right\} e^{-\i (kr_\mathbf{ba} + \mathbf{k}_0 \cdot \mathbf{r}_\mathbf{ba})} ~ .
\end{aligned}
\end{equation}

In solving $\langle \hat{S}^\dagger_\mathbf{a} \rangle$ to first order in $\Gamma_0 t$, we may obtain a closed form for the time-dependence of the multi-atom expectation values $\langle \hat{S}^\dagger_\mathbf{a} \hat{S}_\mathbf{a} \hat{S}^\dagger_\mathbf{b} \rangle$ and $\langle \hat{S}_\mathbf{a} \hat{S}^\dagger_\mathbf{a} \hat{S}^\dagger_\mathbf{b} \rangle$ for $\mathbf{a} \neq \mathbf{b}$
\begin{equation}
\begin{aligned}
    \partial_t \langle \hat{S}^\dagger_\mathbf{a} \hat{S}_\mathbf{a} \hat{S}^\dagger_\mathbf{b} \rangle &\approx \Tr( \hat{S}^\dagger_\mathbf{a} \hat{S}_\mathbf{a} \hat{S}^\dagger_\mathbf{b} \mathcal{L}_1[\hat{\rho}] ) =  \left(\i \Delta\omega_\mathbf{b} - \frac{3}{2} \gamma \right) \langle \hat{S}^\dagger_\mathbf{a} \hat{S}_\mathbf{a} \hat{S}^\dagger_\mathbf{b} \rangle \\
    \partial_t \langle \hat{S}_\mathbf{a} \hat{S}^\dagger_\mathbf{a} \hat{S}^\dagger_\mathbf{b} \rangle &\approx \Tr(\hat{S}_\mathbf{a} \hat{S}^\dagger_\mathbf{a} \hat{S}^\dagger_\mathbf{b} \mathcal{L}_1[\hat{\rho}] ) =  \left(\i \Delta\omega_\mathbf{b} - \frac{1}{2}\gamma \right) \langle \hat{S}_\mathbf{a} \hat{S}^\dagger_\mathbf{a} \hat{S}^\dagger_\mathbf{b} \rangle  + \gamma \langle \hat{S}^\dagger_\mathbf{a} \hat{S}_\mathbf{a} \hat{S}^\dagger_\mathbf{b} \rangle~. \\
\end{aligned}
\end{equation}
We then have
\begin{equation}
\begin{aligned}
    \langle \hat{S}^\dagger_\mathbf{a} \mathcal{F}( t ) \rangle &= e^{-\gamma t / 2} \left\{  e^{\i \Delta\omega_\mathbf{a} t} \left( 1 - \frac{\Gamma_0 t}{ 2 } \right) \langle \hat{S}^\dagger_\mathbf{a} \rangle + \i t \sum_{\mathbf{b}\neq \mathbf{a}} h^{(2)}_\mathbf{ba} e^{\i \Delta\omega_\mathbf{b} t} \left[\left( 2 e^{-\gamma t} - 1 \right) \langle \hat{S}^\dagger_\mathbf{a} \hat{S}_\mathbf{a} \hat{S}^\dagger_\mathbf{b} \rangle - \langle \hat{S}_\mathbf{a} \hat{S}^\dagger_\mathbf{a} \hat{S}^\dagger_\mathbf{b} \rangle \right] \right\} \\
    \langle \hat{S}^\dagger_\mathbf{a} \hat{S}_\mathbf{a} \hat{S}^\dagger_\mathbf{b} \mathcal{F}(t) \rangle &= e^{-3\gamma t/ 2 + \i \Delta\omega_\mathbf{b} t} \langle \hat{S}^\dagger_\mathbf{a} \hat{S}_\mathbf{a} \hat{S}^\dagger_\mathbf{b} \rangle \\
    \langle \hat{S}_\mathbf{a} \hat{S}^\dagger_\mathbf{a} \hat{S}^\dagger_\mathbf{b} \mathcal{F}(t) \rangle &= e^{-\gamma t/ 2 + \i \Delta\omega_\mathbf{b} t} \left[ \left( 1 - e^{-\gamma t} \right) \langle \hat{S}^\dagger_\mathbf{a} \hat{S}_\mathbf{a} \hat{S}^\dagger_\mathbf{b} \rangle  + \langle \hat{S}_\mathbf{a} \hat{S}^\dagger_\mathbf{a} \hat{S}^\dagger_\mathbf{b} \rangle \right) ~ .
\end{aligned}
\end{equation}
Matrix multiplication, dropping terms $\mathcal{O}(\Gamma_0^2 T^2)$, then gives 
\begin{equation}
\begin{aligned}
    \langle \hat{Z}_\mathbf{a} \mathcal{U} \rangle_0 = n_\mathbf{a}\sin(\theta_\mathrm{in}) e^{-\gamma T / 2} \left[  \sin(\phi_\mathrm{out} + \Phi) J_\mathbf{a} + \cos(\phi_\mathrm{out} + \Phi) \left(1 - K_\mathbf{a}\right) \right] 
\end{aligned}
\end{equation}
with
\begin{equation}
\begin{aligned}
    J_\mathbf{a} &= \sum_{\mathbf{b} \neq \mathbf{a}} n_\mathbf{b} \Re\left( V_\mathbf{ba} F_\mathbf{ba}^+\right) \\
        &= \cos(\theta_\mathrm{in}) T \sum_{\mathbf{b} \neq \mathbf{a}} n_\mathbf{b} \Re(V_\mathbf{ba}) + \mathcal{O}(\Gamma_0 \gamma T^2, \Gamma_0 \Delta\omega T^2) \\
    K_\mathbf{a} &= \frac{\Gamma_0 T}{2} + \sum_{\mathbf{b} \neq \mathbf{a}} n_\mathbf{b} \Im\left( V_\mathbf{ba} F_\mathbf{ba}^- \right) \\
        &= \frac{\Gamma_0 T}{2} + \mathcal{O}(\Gamma_0 \gamma T^2, \Gamma_0 \Delta\omega T^2)\\
    F_\mathbf{ba}^\pm &= \frac{T}{4} \boldsymbol{\Bigg(}e^{-\gamma T} \cos\theta_\mathrm{in} - \left(1 - e^{-\gamma T/4}\right)^3 \left(1 + e^{-\gamma T / 4} \right) \\
            &\qquad\quad ~ + \left\{e^{- \gamma T / 4} \left(1 + \cos\theta \right) - 1 \pm 2 \left[ e^{-3 \gamma T / 4} \left(1 + \cos\theta \right) -2 e^{-\gamma T / 2} + 1\right]\right\}e^{\i (\Delta\omega_\mathbf{b} - \Delta\omega_\mathbf{a})T/4}\boldsymbol{\Bigg)}\\
        &= \frac{T}{4} \left\{ e^{-\gamma T} \left( 1 + \cos\theta \right) - 2 e^{-3 \gamma T / 4} + e^{-\gamma T / 4} \left( 3 + \cos\theta \right) - 2 \right. \\
&\qquad\quad \left. \pm 2\left[ e^{-3 \gamma T/4}\left(1 + \cos\theta \right) -2 e^{-\gamma T / 2} + 1\right] + \mathcal{O}(\Delta\omega T) \right\}\\
    \Phi &= \Delta\phi(T) - 2 \Delta\phi(3T/4) + 2 \Delta\phi(T/4) - \Delta\phi(0)
\end{aligned}
\end{equation}
where $\Delta\phi(t)$ is the instantaneous laser phase noise at time $t$.

Averaging over shot-to-shot fluctuations in $\Phi$, which are normally distributed, for $\cos\phi_\mathrm{out} = 0$, the mean $\bar{D}_\mathbf{A}$ and standard deviation $\Delta\bar{D}_\mathbf{A}$ of $\sum_{\mathbf{a} \parallel \mathbf{A}} \langle \hat{Z}_\mathbf{a} \mathcal{U} \rangle_0 / C$ are
\begin{equation}
\begin{aligned}
    \label{eqn:difference-signal}
    \bar{D}_\mathbf{A} &= e^{-\langle \Phi^2 \rangle / 2} \sum_{\mathbf{a} \parallel \mathbf{A}} n_\mathbf{a} J_\mathbf{a} \\
        &= \sum_{\mathbf{a} \parallel \mathbf{A}} n_\mathbf{a} J_\mathbf{a} + \mathcal{O}(\langle \Phi^2 \rangle)\\
    \Delta\bar{D}_\mathbf{A} &= e^{-\langle \Phi^2 \rangle / 2} \sum_{\mathbf{a} \parallel \mathbf{A}} n_\mathbf{a} \sqrt{\left[ \cosh(\langle \Phi^2 \rangle) - 1 \right] J_\mathbf{a}^2 + \sinh(\langle \Phi^2 \rangle) (1 - K_\mathbf{a})^2 } \\
        &= \sqrt{\langle \Phi^2 \rangle} \sum_{\mathbf{a} \parallel \mathbf{A}} n_\mathbf{a}(1 - K_\mathbf{a}) + \mathcal{O}(\langle \Phi^2 \rangle)~.
\end{aligned}
\end{equation}
All modeled quantities presented in the main text use the full time dependencies in Eq.~\ref{eqn:difference-signal} with $\sqrt{ \langle \Phi^2 \rangle } = \sqrt{3 \langle \Delta\phi^2(T) \rangle/ 8} \approx 110~\mathrm{mrad.}$, and $\Delta\omega_\mathbf{a} = \Delta\omega_\mathbf{b} = 0$.
The noise-induced suppression of the mean difference signal $\bar{D}_\mathbf{a}$, $1 - e^{-\langle \Phi \rangle^2/2} \approx 6 \times 10^{-3}$, is insignificant with respect to the measurement precision of the presented data.

\subsection{Finite size effects on collective Lamb Shifts}
\begin{figure}
    \includegraphics[width=3.375in]{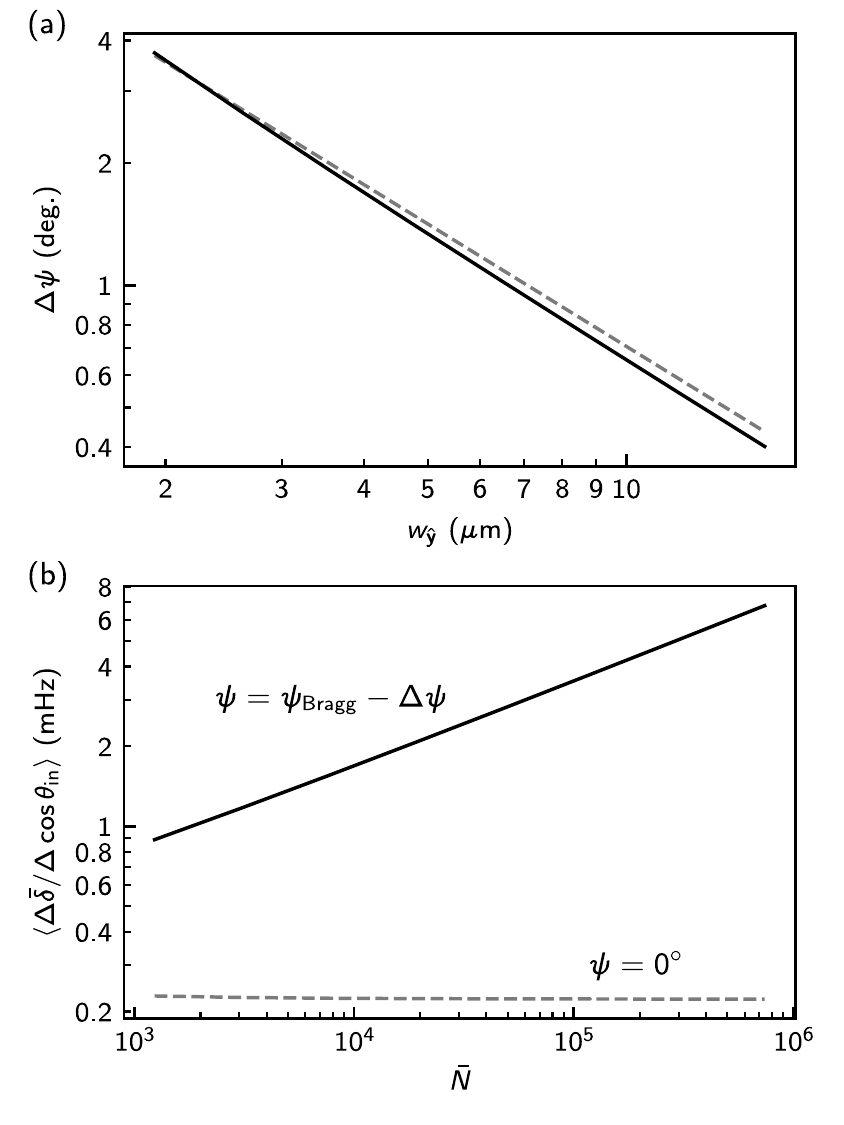}
    \caption{
	Finite size effects on collective Lamb shifts.
	(a) Angular detuning $\Delta \psi = \psi_\mathrm{Bragg} - \psi$ from the Bragg resonance at $\psi_\mathrm{Bragg} \approx 30.8^\circ$ at which the ensemble-averaged collective lamb shift is maximized.
	Values obtained via numerical optimization after global rescaling of the cloud widths $w_{\hat{\mathbf{n}}}$ with fixed $\mu_\mathbf{0} = 1.44$ are shown as the solid black line.
	The diffraction limit $\Delta\psi \approx \pi / 2 k (w_{\hat{\mathbf{x}}} + w_{\hat{\mathbf{y}}})^{-1/2}$ is shown as the grey dashed line.
	(b) Scaling of ensemble-averaged collective Lamb shift sensitivity $\langle \Delta\bar{\delta} / \Delta \cos\theta_\mathrm{in} \rangle$  versus summed counts $\bar{N}$ for $\psi = \psi_\mathrm{Bragg} - \Delta\psi$ (solid black line) and $\psi = 0^\circ$ (grey dashed line).
    }
    \label{fig:finite-size}
\end{figure}
Scalings of the resonant angles of incidence and ensemble-averaged collective Lamb shifts with system size are shown in Fig.~\ref{fig:finite-size}.

\end{document}